\documentclass[]{spie}  %>>> use for US letter paper
\usepackage{amsmath,amsfonts,amssymb}
\usepackage{graphicx}
\usepackage[colorlinks=true, allcolors=blue]{hyperref}

\title{Nancy Grace Roman Space Telescope Coronagraph Instrument Overview and Status}

\author[a]{Vanessa P. Bailey}
\author[a]{Eduardo Bendek}
\author[a]{Brian Monacelli}
\author[a]{Caleb Baker}
\author[a]{Gasia Bedrosian}
\author[a]{Eric Cady}
\author[b]{Ewan S. Douglas}
\author[c]{Tyler Groff}
\author[a]{Sergi R. Hildebrandt}
\author[d]{N. Jeremy {Kasdin}}
\author[a]{John Krist}
\author[e, f]{Bruce Macintosh}
\author[a]{Bertrand Mennesson}
\author[a]{Patrick Morrissey}
\author[a]{Ilya Poberezhskiy}
\author[c]{Hari B. Subedi}
\author[a]{Jason Rhodes}
\author[c]{Aki Roberge}
\author[a]{Marie Ygouf} 
\author[a]{Robert T. Zellem}
\author[a]{Feng Zhao}
\author[c]{Neil T. Zimmerman}

\affil[a]{Jet Propulsion Laboratory, California Institute of Technology, 4800 Oak Grove Drive, Pasadena CA 91109, USA}
\affil[b]{Steward Observatory and the Department of Astronomy, The University of Arizona, 933 N Cherry Ave, Tucson AZ 85719, USA}
\affil[c]{Goddard Space Flight Center, 8800 Greenbelt Rd, Greenbelt MD 20771, USA}
\affil[d]{Department of Mechanical and Aerospace Engineering, Princeton University, 41 Olden St., Princeton NJ 08544, USA}
\affil[e]{University of California Observatories, 1156 High Street, Santa Cruz CA 95064, USA}
\affil[f]{Kavli Institute for Particle Astrophysics and Cosmology, Stanford University, 452 Lomita Mall, Stanford CA 94305, USA}

\authorinfo{Further author information: Send correspondence to V.P.B: vanessa.bailey@jpl.nasa.gov. This document has been reviewed and determined not to contain export controlled technical data. CL\#23-4787}

\begin{document} 
\maketitle

\begin{abstract} 

The Nancy Grace Roman Space Telescope Coronagraph Instrument is a critical technology demonstrator for NASA’s Habitable Worlds Observatory. With a predicted visible-light flux ratio detection limit of $10^{-8}$ or better, it will be capable of reaching new areas of parameter space for both gas giant exoplanets and circumstellar disks.  It is in the final stages of integration and
test at the Jet Propulsion Laboratory, with an anticipated delivery to payload integration in the coming year. This paper will review the instrument systems, observing modes, potential observing applications, and overall progress toward instrument integration and test.

\end{abstract}

\keywords{high-contrast imaging, coronagraphy, Roman Space Telescope, electron multiplying CCDs, deformable mirrors, exoplanets, circumstellar disks}

\section{INTRODUCTION}
\label{sec:intro}  

\begin{figure} [ht]
   \begin{center}
   \begin{tabular}{c} 
   \includegraphics[height=7cm]{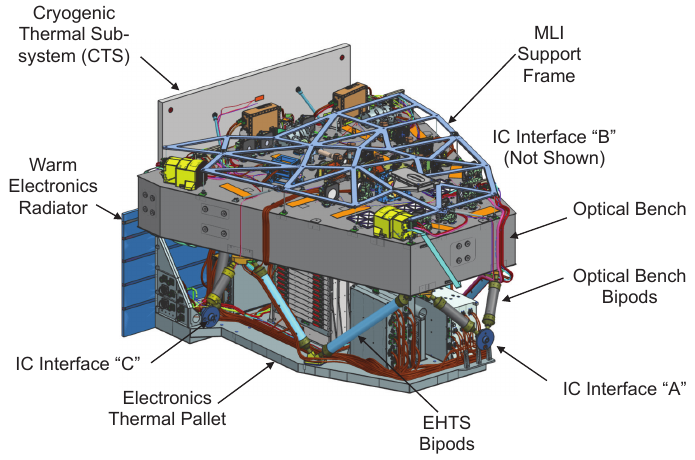} 
   \end{tabular}
   \end{center}
   \caption{The major subsystems of the Coronagraph Instrument\cite{Poberezhskiy2022}. The optical bench sits above the electronics pallet. The Electronics Heat Transport System (EHTS) transports heat from the electronics to the warm radiator. The Cryogenic Thermal Subsystem provides passive cooling to the detectors. Multi-layer insulation (MLI) provides additional thermal shielding. The Instrument Carrier (IC) is the observatory structure to which Coronagraph is mounted.} 
   \label{fig:OBSA+EHTS CAD} 
\end{figure}

\begin{figure} [ht]
   \begin{center}
   \begin{tabular}{c} 
   \includegraphics[height=6cm]{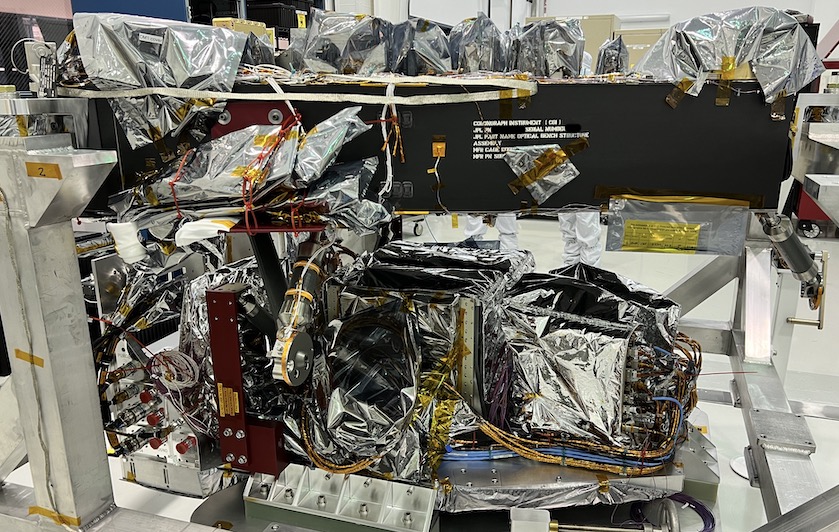}
   \end{tabular}
   \end{center}
   \caption{The optical bench mounted above the EHTS shortly after integration; wiring between the control electronics and optical components had not yet been installed. The optical elements are covered with protective tents for contamination control.} 
   \label{fig:OBSA+EHTS pic} 
\end{figure}

The Nancy Grace Roman Coronagraph Instrument (Figures \ref{fig:OBSA+EHTS CAD} and \ref{fig:OBSA+EHTS pic}) is a technology demonstration instrument that will pave the way for an exo-Earth characterization instrument on NASA's next flagship, the Habitable Worlds Observatory (HWO). Key technologies demonstrated include: high-actuator count deformable mirrors, photon-counting electron-multiplying charge coupled device (EMCCD) detectors, coronagraph masks optimized for an obscured pupil, high-precision high-order wavefront sensing from science camera images, and low-order wavefront sensing and control utilizing starlight rejected by the coronagraph focal plane mask.

\begin{figure} [ht]
   \begin{center}
   \begin{tabular}{c} 
   \includegraphics[height=8cm]{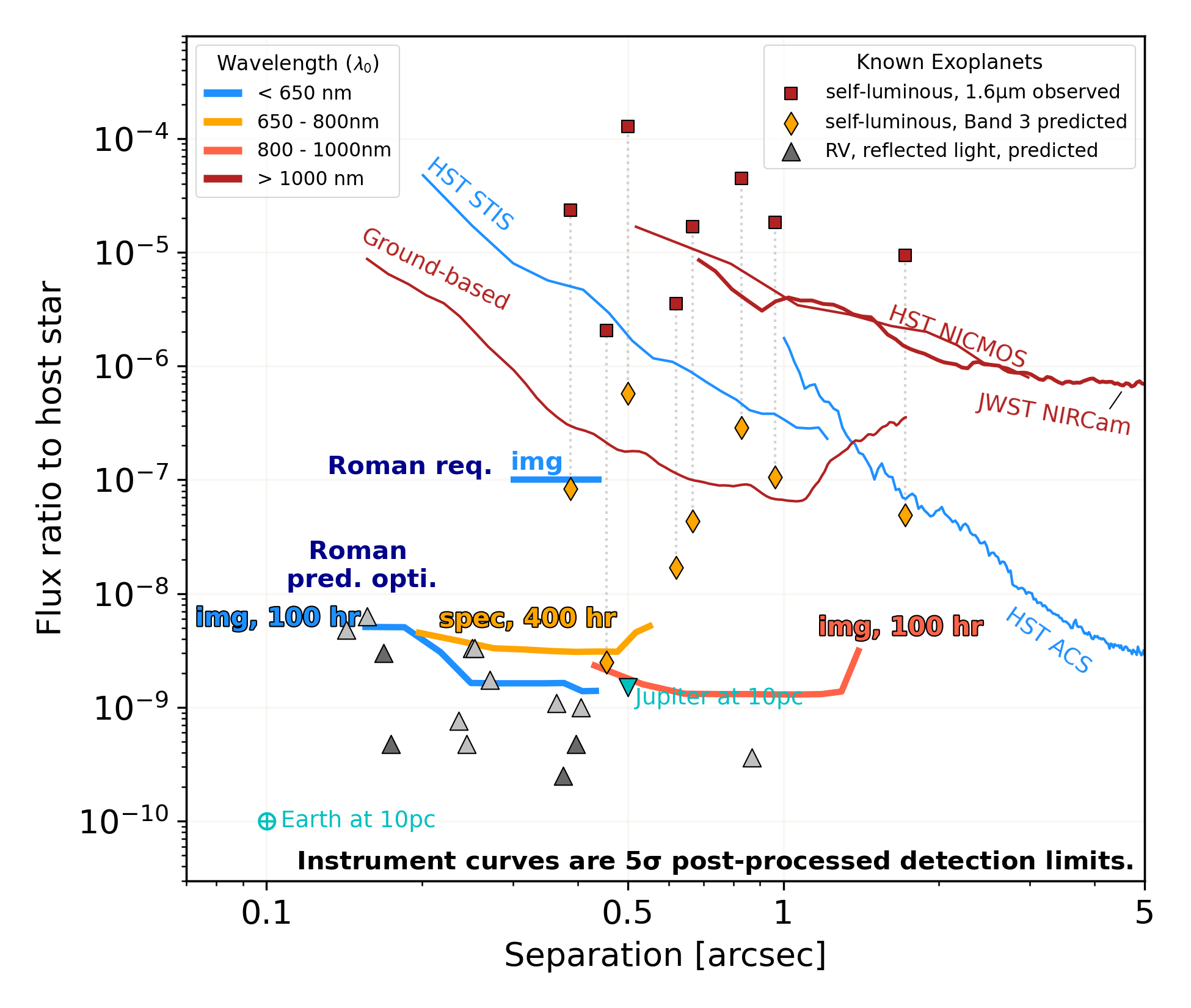}
   \end{tabular}
   \end{center}
   \caption{The Roman Coronagraph Instrument required and predicted detection limits, compared to those of current facilities and to predicted flux ratios for known exoplanets\cite{ThePlot}. Red squares are measured infrared flux ratios of known self-luminous planets; yellow diamonds are the predicted flux ratios for those same planets at Band 3 (730~nm)\cite{Lacy2020}. Grey triangles are predicted flux ratios for known radial velocity planets in reflected visible light at quadrature\cite{IMD, Batalha2018}. Not shown are circumstellar disks, whose flux ratios span from debris disks easily detected by current facilities to faint exozodi disks potentially detectable at the Coronagraph Instrument's predicted performance at the shortest wavelength.} 
   \label{fig:perfplot} 
\end{figure}

With these technologies, the Coronagraph Instrument is predicted to achieve detection limits of better than $10^{-8}$ flux ratio at separations of $3-9~\lambda/D$ ($150-450$~mas at 575~nm). This is several orders of magnitude improvement over current visible light facilities, but still more than an order of magnitude above HWO needs (Figure \ref{fig:perfplot}). Part of the shortfall is because the Roman observatory was not designed nor optimized for coronagraphy. In fact, several of the Roman Coronagraph Instrument subsystems are already predicted to perform in family with the needs articulated in the pre-decadal HabEx and LUVOIR studies\cite{Mennesson2020}. Many important technological challenges remain in HWO development, but the Roman Coronagraph Instrument technology demonstrations will buy down risk.

In this paper, we present short summaries of the key instrument subsystems, observing capabilities, and selected developments from instrument integration and test. Prior proceedings\cite{Poberezhskiy2022, Mennesson2022} provide excellent overviews of the Coronagraph Instrument and its key subsystems and technologies, and we refer the reader to these publications and the others referenced here for further details.

\section{INSTRUMENT}

\subsection{Optical Elements}
\begin{figure} [ht]
   \begin{center}
   \begin{tabular}{r l} 
   \includegraphics[height=6.75cm]{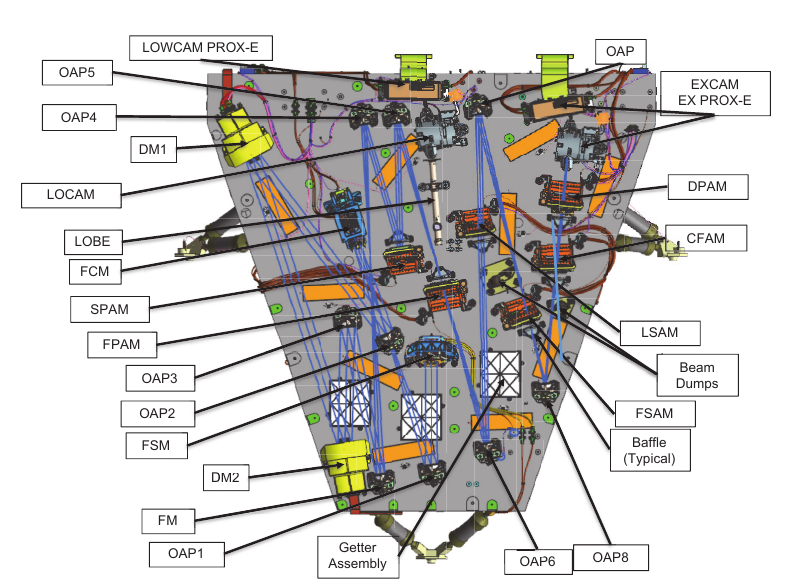} & \includegraphics[height=5.25cm]{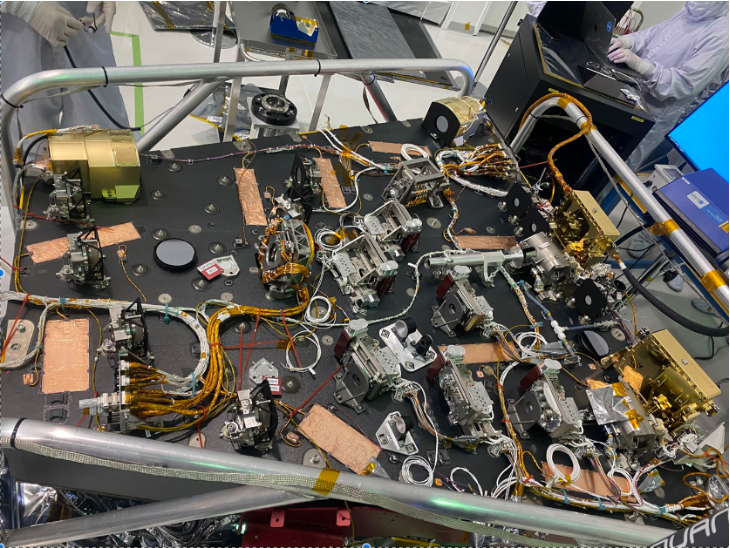}
   \end{tabular}
   \end{center}
   \caption{Optical system diagram (left) and as-built with optical elements uncovered (right).} 
   \label{fig:OBSA} 
\end{figure}

The optical system (Figure \ref{fig:OBSA}) is composed of eight off-axis paraboloidal (OAP) mirrors that form four relay pairs. These optical relays create four different pupil planes and four focal planes along the optical path. Precise alignment and registration of these pupil planes is essential for coronagraph operation in that these are the planes in which filtering is done via the Lyot stop and the shaped-pupil masks (installed in the SPAM: Shaped Pupil Alignment Mechanism). The Fast Steering Mirror (FSM) and first Deformable Mirror (DM1) are also located in pupil planes to manipulate the wavefront exactly in its pupil. The second deformable mirror (DM2) is at an intermediate plane to correct amplitude errors. The Focus Control Mechanism (FCM) provides control of slow focus drift. Two of the focal points are used for coronagraph operation: one allows the focal-plane alignment mechanism (FPAM) to insert an occulting mask or to otherwise filter the beam and direct light to the low-order wavefront sensor camera (LOCAM); the second focus allows for insertion of different field stop sizes and shapes via the Field Stop Alignment Mechanism (FSAM). The Color Filter Alignment Mechanism (CFAM) holds the various engineering and observation color filters near a pupil. The Dispersion Polarization Alignment Mechanism (DPAM) holds the the direct imaging lens, pupil lens, phase-retrieval lenses, zero-deviation prisms, and two Wollaston prisms. Finally, the image is focused on the Exoplanetary Systems Camera (EXCAM). Please refer to references \citenum{Poberezhskiy2022, Tang2019, Groff2021, Groff2023, Caillat2022} for more details. 

\subsection{Detectors}
\label{sec:detectors}

\begin{figure} [ht]
   \begin{center}
   \begin{tabular}{c} 
   \includegraphics[height=6cm]{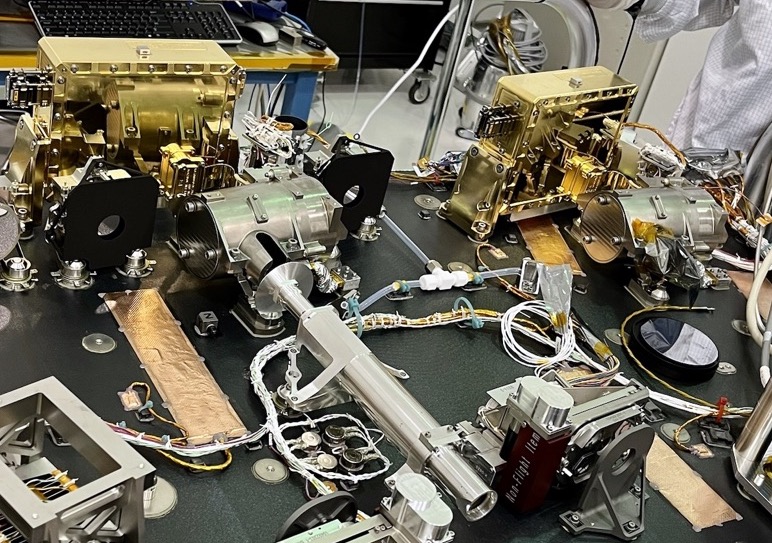} 
   \end{tabular}
   \end{center}
   \caption{Detectors installed on the optical bench. The two gold boxes are the camera proximity electronics; the silver cylindrical enclosures in front of them house the detectors. LOCAM is on the left and EXCAM is on the right.} 
   \label{fig:EMCCDs-bench} 
\end{figure}

\begin{figure} [ht]
   \begin{center}
   \begin{tabular}{c} 
   \includegraphics[height=4.75cm]{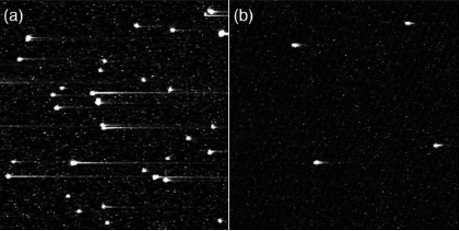} 
   \end{tabular}
   \end{center}
   \caption{Improvements in cosmic ray ``tails'' vs. commercial CCDs\cite{Morrissey2023}. An overspill for the gain register reduces the length of cosmic ray tails in Roman's EMCCDs (b) by more than 2x relative to commercial devices (a). } 
   \label{fig:EMCCDs-CR} 
\end{figure}

The Coronagraph Instrument has two cameras: the Low Order Camera (LOCAM) and Exoplanetary Systems Camera (EXCAM). Figure \ref{fig:EMCCDs-bench} shows the two cameras and their proximity electronics installed on the optical bench. Significant development was invested in the radiation hardness and cosmic ray resilience (Figure \ref{fig:EMCCDs-CR}). Together, the various improvements have yielded a factor of $\sim3$ improvement in effective quantum efficiency. We refer the reader to \citenum{Daigle2022, Morrissey2023, Bush2023} for a thorough discussion.

\subsection{Deformable Mirrors}
Two Xinetics 48x48 actuator deformable mirrors (DMs) provide wavefront control for all modes except tip, tilt, and focus, with mean stroke resolution of 7.5~pm. Reference \citenum{Krist2023} provides information about the as-built actuator yield, influence functions, and long-term creep, as well as the observed astigmatism on both mirrors both before and after mitigation by static optics realignment.

\subsection{Coronagraph Masks}
The large central obscuration and six wide secondary mirror supports create a challenging pupil for coronagraphy. Both (Hybrid) Lyot and Shaped Pupil masks are installed for starlight suppression. Additionally, a transmissive Zernike wavefront sensor mask is installed, though not currently supported. The flight mask designs and their design considerations are presented in \citenum{Riggs2021} and \citenum{Gersh-Range2022}.

\subsection{Prisms}
In addition to integrated light imaging, the Coronagraph Instrument has ``best effort'' low-resolution ($R\sim50$) single slit spectroscopy and modest-precision polarimetry (3\% RMSE) modes\cite{Mennesson2022} (Table \ref{tab:modes}). These capabilities are enabled by a zero-deviation spectroscopy prism and a pair of Wollason prisms.\cite{Groff2021, Groff2023}.

\subsection{Wavefront Sensing and Control}
Low Order Wavefront Sensing (LOWFS) is accomplished with a dedicated sensor that uses light rejected by the coronagraph focal plane masks. High speed tip/tilt control (1kHz framerate; $<1$~mas residual) is essential to achieving the design level of starlight suppression. Low speed  Z4 - Z11 (of order mHz bandwidth; $<10$~pm residual) is also available\cite{Krist2023, Shi2016, Shi2018, Dube2022}. In Z4 - Z11, Roman's LOWFS/C system outperforms the requirements given in the LUVOIR and HabEx reports\cite{Mennesson2020}.

High Order Wavefront Sensing (HOWFS) uses EXCAM images, rather than a dedicated wavefront sensor, to minimize non-common path aberrations\cite{Krist2023, Zhou2023}. HOWFSC is implemented ``ground in the loop'' (GITL), because onboard processing power is extremely limited. HOWFS images are downlinked and automatically analyzed in real time to compute the next iteration of DM commands, which are then uplinked.  A GITL architecture \textit{in principle} enables more flexibility to update the WFSC algorithm after launch, and several groups are researching alternatives (eg: \citenum{Sirbu2023, Milani2023} in this conference). Whether updates are possible in practice will depend on testbed resources available after the Coronagraph Instrument has been delivered.

\subsection{Integrated Modeling and Error Budgeting}
\label{sec:IM+OS}
Maturation and validation of integrated modeling and analytic error budgets are in many ways as important for HWO as any of the aforementioned technologies. In-depth discussions of the system model and modeling approaches are presented in \citenum{Krist2023, Krist2023SPIE, Nemati2023, Zhou2023}, and references therein. Simulated speckle field images from these ``Observing Scenarios'' are publicly available\footnote{\url{https://roman.ipac.caltech.edu/sims/Coronagraph_public_images.html}}\cite{Krist2023}.

\section{OBSERVING CAPABILITIES}

The launch date commitment is May 2027, with a current planned date of October 2026. The nominal time allocations for the Coronagraph Instrument remain unchanged from previous reports: 450~hr spread across the initial commissioning phase (launch + 3 months) and 2200~hr spread across multiple observing campaigns during the so-called ``Tech Demo Phase'' (subsequent 18 months). Additional time beyond the Tech Demo Phase could potentially be granted if performance is sufficiently compelling. The Roman mission reserves 25\% of its time for guest investigator programs (on the Wide Field Instrument); additional Coronagraph Instrument time would count against this pool and is not guaranteed.

\subsection{Observing Modes}
\label{sec:modes}
The only formal performance requirement for the Coronagraph Instrument is to be able to image a point source with a flux ratio of at least $10^{-7}$ at a separation of $6-9~\lambda/D$ from a star as faint as $V=5$ in a $10\%$ bandpass centered $<600$~nm. As such, there is only one required, and hence fully-supported, observing mode. However, additional coronagraph masks, color filters, and spectroscopy and polarimetry prisms are installed in the instrument (Table \ref{tab:modes}). Furthermore, in many modes, the Coronagraph Instrument should be capable of achieving detection limits well beyond its $10^{-7}$ requirement (Figure \ref{fig:perfplot}).

\begin{table}[htb]
  \caption{Supported and ``best effort'' observing modes list. Additional currently-unsupported modes, ranging from additional filter + field of view combinations to alternative WFSC-enabling masks, are installed but will not be performance-tested prior to instrument delivery. A description of the full complement of flight masks is available in \citenum{Riggs2021}. **Polarimetry with Band 1 imaging is ``best effort.''}
  \label{tab:modes}
  \includegraphics[width=\linewidth]{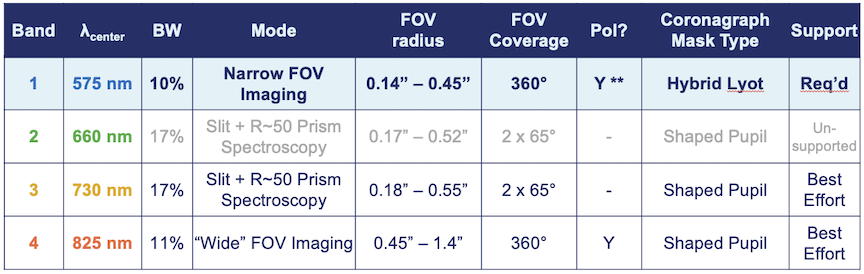}
\end{table}

\subsection{Potential Observation Applications}
\label{sec:definitely not science}

The top priority observation will be to demonstrate the Coronagraph Instrument's performance requirement, described in Section \ref{sec:modes}. Beyond that, as resources are available, we will attempt to commission one or more ``best effort'' or ``unsupported'' modes. Potential observation applications include, but are not limited to:

\begin{itemize}
    \item using the required HLC imaging mode: Debris disk imaging at complementary wavelengths and spatial scales to ground, HST, and JWST \cite{Milani2021}
    \item using the required HLC imaging mode at goal performance:
        \begin{itemize}
        \item{Potential for the first visible light image of an exozodiacal disk; exozodi survey\cite{Douglas2022}}
        \item{Potential for the first reflected light image of a cold Jupiter analog\cite{Batalha2018}}
        \end{itemize}
    \item using the ``best effort'' polarimetry modes: Debris disk polarimetry\cite{Anche2022, Anche2023, Doelman2023}
    \item using the ``best effort'' spectroscopy mode: 
    \begin{itemize}
        \item{Spectra of self-luminous planets at wavelengths complementary to ground, HST, and JWST\cite{Lacy2020}}
        \item{Potential for the first visible light spectrum of a cold Jupiter analog\cite{Saxena2021}}
        \end{itemize}
    \item{using the currently-unsupported multistar mask: high contrast imaging in binary systems\cite{Sirbu2023}}
    \item{using the currently-unsupported transmissive Zernike mask: high-precision wavefront monitoring\cite{Ruane2020}}
\end{itemize}

\section{INTEGRATION AND TEST}

\subsection{Optical Alignment}

The optics of the Coronagraph Instrument were interferometrically aligned in the backwards direction to best match the interferometer’s 9~mm beam diameter to the Coronagraph Instrument's 5-mm exit pupil diameter, and to best fit the significant amount of optical and optomechanical hardware on the grand piano-sized optical bench. 
 
Each mirror was mounted in its own assembly, and mirror poses were controlled via flexurized Moore rods. Adjustment of these rods allowed precise alignment of each mirror’s orientation in azimuth, elevation, and clocking angles. The lateral and focus alignment of each assembly was adjustable via eccentric bushings (in the plane of the optical bench), and mirror height was aligned via shims. Each mirror’s optical assembly was initially located on the optical bench via coordinate measurement machine (CMM) arm. This CMM arm was used to position optical references to align sequentially each OAP mirror, whether via a pupil relay optical system located at a OAP mirror focus, or a flat mirror located at the relayed pupil after each OAP mirror relay pair. 
 
End-to-end optical alignment was first achieved using flat surrogate mirrors for each of the active mechanisms (including the DMs, FSM, FCM, and SPAM mirror). With the surrogate flats, end-to-end wavefront error across the Coronagraph Instrument pupil was $<15$~nm RMS against a 60~nm RMS requirement. After insertion of the flight mechanisms, the wavefront error was maintained until the DMs were installed. DM-native aberrations (primarily astigmatism due to dry-out) increased the end-to-end wavefront error significantly, so some of the OAP mirrors were deliberately misaligned to compensate as much as possible for these aberrations. The net result was an end-to-end wavefront error measurement of $<45$~nm RMS that maintained the registration of all pupils and preserved the boresight angle into the instrument. 

\subsection{The Coronagraph Verification Stimulus}
% this section was written primarily by Eduardo Bendek. 

The Coronagraph Verification Stimulus (CVS) is an Optical Ground Support instrument that will be used to verify the Coronagraph Instrument in air and in vacuum. CVS will deliver a beam of light that emulates the input from the telescope, including flight-like jitter perturbations. CVS is held above the Coronagraph Instrument optical bench to deliver the telescope pupil on the FSM. The CVS is held by a Support Structure which is attached to the Thermal Vacuum (TVAC) fixture. The system is fed by a supercontinuum laser (Super K Extreme) and wavelength selection system called VARIA. The laser output is coupled to a Photonic Crystal Fiber (PCF) capable of transmitting the observation wavelength range ($\sim525 - 840$~nm) in single mode. The fiber injects the light in a $3~\mu$m pinhole to produce a single source equivalent. Then, light is collimated by a 60'' focal length OAP working at f/38.1. Downstream of the collimating OAP is located the system stop, which is a 40~mm aperture version of the Phase-C Roman pupil. Then, a pair of OAPs demagnify the pupil to 20~mm, reimaging it on the surface of the Jitter Mirror (JM), which provides the ability to inject Z2 and Z3 (tip/tilt) perturbations. From there, another set of OAPs magnifies the pupil and re-images it at the CVS Exit Pupil which should match the FSM location. In the optical path, before reaching the exit pupil, a periscope can tilt and translate the pupil.   

\begin{figure} [ht]
   \begin{center}
   \begin{tabular}{c} 
   \includegraphics[height=8cm]{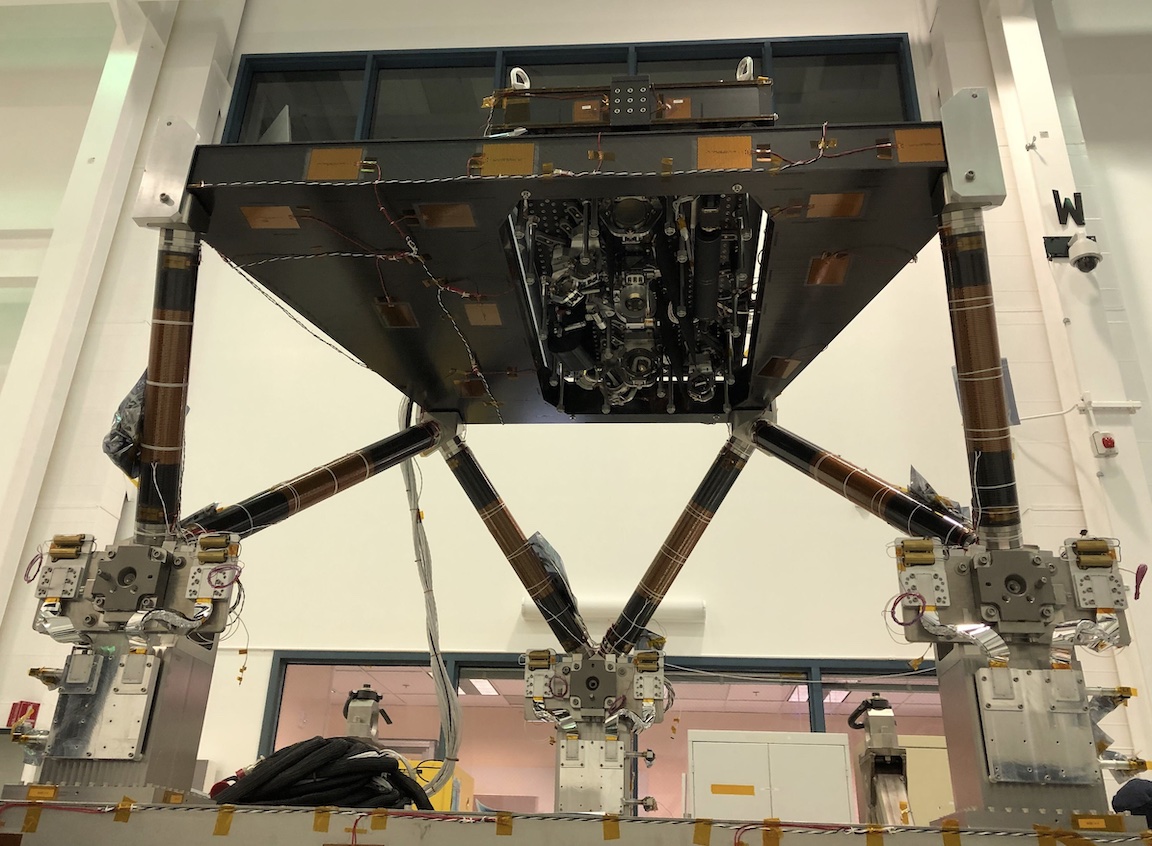}
   \end{tabular}
   \end{center}
   \caption{View of the CVS optical bench held by the Support Structure. The Coronagraph Instrument will be placed in the space below CVS during pre-delivery testing.} 
   \label{fig:CVS+SS} 
\end{figure}

\subsection{Upcoming Integration and Test Activities}

Two instrument-level test campaigns remain before instrument delivery to payload integration and test, nominally May 2024. Full functional test (FFT) in the fall of this year will exercise basic instrument functionality in air. Thermal Vacuum (TVAC) tests will test performance in a stable flight-like pressure and thermal environment, including cooled detectors. Dark hole performance in TVAC will exceed that of FFT but will not reach that predicted to be achieved on-sky. One reason is the inherent stability limit of the test chamber and CVS themselves. Another is that the settling time for DM actuator creep\cite{Krist2023} is long compared to the duration of the TVAC high-performance campaign. Refer to \citenum{Poberezhskiy2022} for a more complete description of instrument integration and test plans. 

During payload integration, instrument-to-observatory alignment checks will be executed, but dark hole digging will not. The gravity sag exhibited by the primary mirror introduces wavefront errors much larger than the Coronagraph Instrument is designed to correct, so point source imaging and coronagraphy tests would not produce meaningful results.

\section{OTHER RESOURCES}
Curated collections of resources on the Roman Mission and the Coronagraph Instrument are available online\footnote{\url{https://roman.gsfc.nasa.gov/science/roses.html}}\footnote{\url{https://roman.ipac.caltech.edu/SSC_Supplemental_info_for_Roses_call.html}}. Users interested in image simulation and analysis may be interested in the Observing Scenarios described in Section \ref{sec:IM+OS}, their post-processing analysis reports\cite{YgoufOS9HLC, YgoufOS9SPC, Krist2023}, and the Roman Coronagraph Instrument Data Challenge.\footnote{\url{https://www.exoplanetdatachallenge.com/}.} A wide variety of open-source image, WFSC, target, and mission simulation tools are available\cite{Douglas2020}.

\section{SUMMARY}

The Roman Coronagraph Instrument, a technology demonstrator for NASA's future Habitable Worlds Observatory flagship, is predicted to achieve unprecedented starlight suppression in visible wavelengths. Although it has only one required observing mode, a suite of installed coronagraph masks, spectroscopy prisms, and polarimeters are also installed, providing latent capability for additional compelling technology demonstrations and exoplanetary system science applications.  It is in the final stages of integration and test at the Jet Propulsion Laboratory, with an anticipated delivery to payload integration of May 2024. 

\iffalse %bulk comment

    \appendix    %>>>> this command starts appendixes
    
    \section{TEMPLATES}
    
        \begin{table}[ht]
        \caption{caption.} 
        \label{tab:example}
        \begin{center}       
        \begin{tabular}{|l|l|l|} 
        \hline
        \rule[-1ex]{0pt}{3.5ex}  Margin & A4 & Letter  \\
        \hline
        \rule[-1ex]{0pt}{3.5ex}  Top margin & 2.54 cm & 1.0 in.   \\
        \hline 
        \end{tabular}
        \end{center}
        \end{table}
    
    An example equation 
        \begin{equation}
        \label{eq:fov}
        2 a = \frac{(b + 1)}{3c}. 
        \end{equation}
    
\fi

\acknowledgments % equivalent to \section*{ACKNOWLEDGMENTS}       
 
The research was carried out in part at the Jet Propulsion Laboratory, California Institute of Technology, under a contract with the National Aeronautics and Space Administration (80NM0018D0004).
This research has made use of the Imaging Mission Database, which is operated by the Space Imaging and Optical Systems Lab at Cornell University. The database includes content from the NASA Exoplanet Archive, which is operated by the California Institute of Technology, under contract with the National Aeronautics and Space Administration under the Exoplanet Exploration Program, and from the SIMBAD database, operated at CDS, Strasbourg, France.

% References
\bibliography{references} % bibliography data in report.bib

\begin{thebibliography}{10}

\bibitem{Mennesson2020}
Mennesson, B., Juanola-Parramon, R., Nemati, B., Ruane, G., Bailey, V.~P.,
  Bolcar, M., Martin, S., Zimmerman, N., Stark, C., Pueyo, L., Benford, D.,
  Cady, E., Crill, B., Douglas, E., Gaudi, B.~S., Kasdin, J., Kern, B., Krist,
  J., Kruk, J., Luchik, T., Macintosh, B., Mandell, A., Mawet, D.,
  Poberezhskiy, I., Rhodes, J., Riggs, A.~J., Turnbull, M., Roberge, A., Shi,
  F., Siegler, N., Stapelfeldt, K., Ygouf, M., Zellem, R., and Zhao, F.,
  ``Paving the way to future missions: The roman space telescope coronagraph
  technology demonstration,'' {\em arXiv}  (2020).

\bibitem{Poberezhskiy2022}
Poberezhskiy, I.~Y., Heydorff, K., Luchik, T., Zhao, F., Cady, E., Bedrosian,
  G., Colavita, M., Creager, B., Goullioud, R., Groff, T.~D., Grue, A.,
  Monacelli, B., Morrissey, P., Kempenaar, J., Kern, B.~D., King, M.~E., Koch,
  T., Krist, J.~E., Kuan, G.~M., Lam, J.~C., Lewis, D., Mok, F., Muliere, D.,
  Nemati, B., Noecker, C., Oseas, J.~M., Riggs, A.~J., Shi, F., and
  Shreckengost, B., ``Roman coronagraph instrument: engineering overview and
  status,'' {\em Proc. SPIE 12180 Space Telescopes and Instrument 2022:
  Optical, Infrared, and Millimeter Wave}~{\bf 12180},  121801X (2022).

\bibitem{Mennesson2022}
Mennesson, B., Bailey, V.~P., Zellem, R., Hildebrandt, S.~R., Ygouf, M.,
  Rhodes, J., Zimmerman, N.~T., Nemati, B., Gonzalez, G., Cady, E., Kern,
  B.~D., Koch, T., Krist, J., Heydorff, K., Luchik, T., Mok, F., Morrissey, P.,
  Poberezhskiy, I., Riggs, A. J.~E., Shi, F., Zhao, F., Akeson, R., Armus, L.,
  Greenbaum, A., Ingalls, J., and Lowrance, P., ``{The Roman Space Telescope
  coronagraph technology demonstration: current status and relevance to future
  missions},'' {\em Proc. SPIE Space Telescopes and Instrument 2022: Optical,
  Infrared, and Millimeter Wave}~{\bf 12180},  121801W (2022).

\bibitem{ThePlot}
\url{https://github.com/nasavbailey/DI-flux-ratio-plot/}.

\bibitem{Lacy2020}
Lacy, B. and Burrows, A., ``Prospects for directly imaging young giant planets
  at optical wavelengths,'' {\em The Astrophysical Journal}~{\bf 892},  151
  (2020).

\bibitem{IMD}
\url{https://plandb.sioslab.com/about.php}.

\bibitem{Batalha2018}
Batalha, N.~E., Smith, A. J. R.~W., Lewis, N.~K., Marley, M.~S., Fortney,
  J.~J., and Macintosh, B., ``Color classification of extrasolar giant planets:
  Prospects and cautions,'' {\em The Astronomical Journal}~{\bf 156},  158
  (2018).

\bibitem{Tang2019}
Tang, H., Rodgers, M., Creager, B., Krist, J.~E., McGuire, J., Patterson, K.,
  Rud, M., Shi, F., and Zhao, F., ``The wfirst coronagraph instrument phase b
  optical design,'' {\em Proceedings of SPIE - Techniques and Instrumentation
  for Detection of Exoplanets IX}~{\bf 11117},  111170C (2019).

\bibitem{Groff2021}
Groff, T.~D., Zimmerman, N.~T., Subedi, H.~B., Rizzo, M., Titus, J., Lyons, J.,
  Bell, D., Gaylin, S., Gao, G., Pasquale, B.~A., Nicolaeff, N., Tamura, M.,
  and Shi, F., ``Roman space telescope cgi: prism and polarizer
  characterization modes,'' {\em Proc. SPIE Space Telescopes and
  Instrumentation 2020: Optical, Infrared, and Millimeter Wave}~{\bf 11443},
  114433D (2021).

\bibitem{Groff2023}
Groff, T.~D., Zimmerman, N., Subedi, H., Titus, J., Bell, D., Gao, G., Gaylin,
  S., Nicolaeff, N., Yamada, T., and Tamura, M., ``Rst cgi: final verification
  and calibration of prism and polarizer flight units,'' {\em Proc. SPIE
  Techniques and Instrumentation for Detection of Exoplanets XI}~{\bf 12680}
  (2023).

\bibitem{Caillat2022}
{Caillat}, A., {Ferrari}, M., {Marcos}, M., {Pari{\`e}s}, C., {Bonnefoi}, A.,
  {Vitrac}, V., {Roulet}, M., and {Hugot}, E., ``{Super polished mirrors for
  the Roman Space Telescope CoronaGraphic Instrument: design, manufacturing,
  and optical performances},'' {\em Proc. SPIE Space Telescopes and
  Instrumentation 2022: Optical, Infrared, and Millimeter Wave}~{\bf 12180},
  121801Y (2022).

\bibitem{Daigle2022}
Daigle, O., Veilleux, J., Grandmont, F.~J., Morrissey, P., Basset, C., Bush,
  N., Hoenk, M., Gilbert, A., Turcotte, J., and Ghodoussi, A., ``{Nancy Grace
  Roman Space Telescope coronagraph EMCCD flight camera electronics
  development},'' {\em Proc. SPIE Space Telescopes and Instrumentation 2022:
  Optical, Infrared, and Millimeter Wave}~{\bf 12180},  1218064 (2022).

\bibitem{Morrissey2023}
Morrissey, P., Harding, L., Bush, N., Bottom, M., Nemati, B., Daniel, A., Jun,
  B., Martinez, L. M.~S., Desai, N., Barry, D., Davis, R.-T., Demers, R., Hall,
  D., Holland, A., Turner, P., and Shortt, B., ``Flight photon counting
  electron multiplying charge coupled device development for the roman space
  telescope coronagraph instrument,'' {\em JATIS}~{\bf 9} (2023).

\bibitem{Bush2023}
Bush, N., Morrissey, P., Hoenk, M., Kyne, G., Basset, C., Lamborn, A., Letona,
  J., Chen, W., and Nemati, B., ``The roman cgi camera systems: Excam and
  locam,'' {\em JATIS}~{\bf submitted}.

\bibitem{Krist2023}
Krist, J., Steeves, J.~B., Dube, B.~D., Riggs, A. J.~E., Kern, B.~D., Marx,
  D.~S., Cady, E.~J., Zhou, H., Poberezhskiy, I.~Y., Baker, C.~W., McGuire,
  J.~P., Nemati, B., Kuan, G.~M., Mennesson, B., Trauger, J.~T., Saini, N.~S.,
  and Rafels, S.~H., ``End-to-end numerical modeling of the roman space
  telescope coronagraph,'' {\em JATIS}  (submitted).

\bibitem{Riggs2021}
Riggs, A.~E., Moody, D.~C., Gersh-Range, J., Sirbu, D., Belikov, R., Bendek,
  E., Bailey, V.~P., Balasubramanian, K., Wilson, D.~W., Basinger, S.~A.,
  Debes, J., Groff, T.~D., Kasdin, N.~J., Mennesson, B., Moore, D.~M., Ruane,
  G., Sidick, E., Siegler, N., Trauger, J., and Zimmerman, N.~T., ``Flight mask
  designs of the roman space telescope coronagraph instrument,'' {\em Proc.
  SPIE Techniques and Instrumentation for Detection of Exoplanets X}~{\bf
  11823},  118231Y (2021).

\bibitem{Gersh-Range2022}
{Gersh-Range}, J., {Riggs}, A.~J.~E., and {Kasdin}, N.~J., ``{Flight designs
  and pupil error mitigation for the bowtie shaped pupil coronagraph on the
  Nancy Grace Roman Space Telescope},'' {\em JATIS}~{\bf 8},  025003 (Apr.
  2022).

\bibitem{Shi2016}
Shi, F., Balasubramanian, K., Hein, R., Lam, R., Moore, D., Moore, J.,
  Patterson, K., Poberezhskiy, I., Shields, J., Sidick, E., Tang, H., Truong,
  T., Wallace, J.~K., Wang, X., and Wilson, D., ``Low-order wavefront sensing
  and control for wfirst-afta coronagraph,'' {\em JATIS}~{\bf 2},  011021
  (2016).

\bibitem{Shi2018}
Shi, F., Kern, B.~D., Marx, D.~S., Patterson, K., Prada, C.~M., Seo, B.-J.,
  Shelton, J.~C., Shields, J., Tang, H., Truong, T., Shaw, J., Cady, E.~J., and
  Lam, R., ``Wfirst low order wavefront sensing and control dynamic testbed
  performance under the flight like photon flux,'' {\em Space Telescopes and
  Instrumentation 2018: Optical, Infrared, and Millimeter Wave}~{\bf 10698},
  106982O (2018).

\bibitem{Dube2022}
Dube, B.~D., Riggs, A.~J., Kern, B.~D., Cady, E.~J., Krist, J.~E., Zhou, H.,
  Nemati, B., Seo, B.-J., Steeves, J., Arndt, D., Mandić, M., Shields, J.,
  Boussalis, D., Valverde, A., Rahman, Z., and Fathpour, N., ``Exascale
  integrated modeling of low-order wavefront sensing and control for the roman
  coronagraph instrument,'' {\em Journal of the Optical Society of America
  A}~{\bf 39},  C133 (12 2022).

\bibitem{Zhou2023}
Zhou, H., Kern, B.~D., Bush, N., Krist, J.~E., and Poberezhskiy, I.~Y., ``Roman
  coronagraph howfsc modeling: case study and raw contrast performance
  prediction,'' {\em Proc. SPIE Techniques and Instrumentation for Detection of
  Exoplanets XI}~{\bf 12680} (2023).

\bibitem{Sirbu2023}
Sirbu, D., Belikov, R., Bendek, E., Marx, D., {Mejia Prada}, C., Riggs, A.
  J.~E., and Zhou, H., ``Multi-star wavefront control at the occulting mask
  coronagraph testbed: monochromatic laboratory demonstration for the roman
  coronagraph instrument,'' {\em Proc. SPIE Techniques and Instrumentation for
  Detection of Exoplanets XI}~{\bf 12680} (2023).

\bibitem{Milani2023}
Milani, K., Douglas, E.~S., Haffert, S., and van Gorkom, K., ``Simulating the
  efficacy of the implicit-electric-field-conjugation algorithm for the roman
  coronagraph with noise,'' {\em Proc. SPIE Techniques and Instrumentation for
  Detection of Exoplanets XI}~{\bf 12680} (2023).

\bibitem{Krist2023SPIE}
Krist, J.~E., Steeves, J., Saini, N., Dube, B., and Arndt, D., ``End-to-end
  numerical modeling of the roman space telescope coronagraphic instrument,''
  {\em Proc. SPIE Techniques and Instrumentation for Detection of Exoplanets
  XI}~{\bf 12680} (2023).

\bibitem{Nemati2023}
Nemati, B. and {et. al}, ``The analytical performance model and error budget
  for the roman coronagraph,'' {\em JATIS}~{\bf submitted}.

\bibitem{Milani2021}
Milani, K. and Douglas, E.~S., ``Faster imaging simulation through complex
  systems: a coronagraphic example,'' {\em Proc SPIE Space Telescopes and
  Instrumentation 2020: Optical, Infrared, and Millimeter Wave}~{\bf 11443},
  1144338 (2021).

\bibitem{Douglas2022}
{Douglas}, E.~S., {Debes}, J., {Mennesson}, B., {Nemati}, B., {Ashcraft}, J.,
  {Ren}, B., {Stapelfeldt}, K.~R., {Savransky}, D., {Lewis}, N.~K., and
  {Macintosh}, B., ``{Sensitivity of the Roman Coronagraph Instrument to
  Exozodiacal Dust},'' {\em PASP}~{\bf 134},  024402 (Feb. 2022).

\bibitem{Anche2022}
Anche, R.~M., Douglas, E.~S., Milani, K., Ashcraft, J., and Debes, J.,
  ``Simulations of polarimetric observations of debris disks through the roman
  coronagraph instrument,'' {\em Proc. SPIE Space Telescopes and Instrument
  2022: Optical, Infrared, and Millimeter Wave}~{\bf 12180},  1218056 (2022).

\bibitem{Anche2023}
Anche, R.~M., Douglas, E., Milani, K., Ashcraft, J., Millar-Blanchaer, M.~A.,
  Debes, J.~H., and Milli, J., ``Measuring the power of polarimetry: simulation
  of debris disk observations with the roman coronagraph,'' {\em PASP}~{\bf
  submitted}.

\bibitem{Doelman2023}
Doelman, D.~S., Belaouchi, H., Riggs, A.~E., Mennesson, B., Ouellet, M.,
  Rietjens, J., Hoevers, H., and Snik, F., ``Falco simulations of high-contrast
  polarimetry with the nancy grace roman space telescope coronagraph
  instrument,'' {\em Proc. SPIE Techniques and Instrumentation for Detection of
  Exoplanets XI}~{\bf 12680} (2023).

\bibitem{Saxena2021}
{Saxena}, P., {Villanueva}, G.~L., {Zimmerman}, N.~T., {Mandell}, A.~M., and
  {Smith}, A. J.~R.~W., ``{Simulating Reflected Light Exoplanet Spectra of the
  Promising Direct Imaging Target, Andromedae d, with a New, Fast Sampling
  Method Using the Planetary Spectrum Generator},'' {\em AJ}~{\bf 162}(1),  30
  (2021).

\bibitem{Ruane2020}
Ruane, G., Wallace, J.~K., Steeves, J., {Mejia Prada}, C., Seo, B.-J., Bendek,
  E., Coker, C., Chen, P., Crill, B., Jewell, J., Kern, B., Marx, D., Poon,
  P.~K., Redding, D., Riggs, A. J.~E., Siegler, N., and Zimmer, R., ``Wavefront
  sensing and control in space-based coronagraph instruments using zernike’s
  phase-contrast method,'' {\em JATIS}~{\bf 6},  045005 (2020).

\bibitem{YgoufOS9HLC}
Ygouf, M., Zimmerman, N., Bailey, V., Krist, J., Zellem, R., and Debes, J.,
  ``Roman coronagraph instrument post processing report - os9 hlc
  distribution,'' tech. rep. (2021).

\bibitem{YgoufOS9SPC}
Ygouf, M., Zimmerman, N., and Bailey, V., ``Roman coronagraph instrument post
  processing report - os9 spc distribution,'' tech. rep. (2022).

\bibitem{Douglas2020}
Douglas, E.~S., Ashcraft, J.~N., Belikov, R., Debes, J., Kasdin, J., Krist, J.,
  Lacy, B., Nemati, B., Milani, K., Pogorelyuk, L., Riggs, A., Savransky, D.,
  and Sirbu, D., ``A review of simulation and performance modeling tools for
  the roman coronagraph instrument,'' {\em Proceedings of SPIE, Space
  Telescopes and Instrumentation 2020: Optical, Infrared, and Millimeter
  Wave}~{\bf 11443},  1144338 (2020).

\end{thebibliography}
\bibliographystyle{spiebib} % makes bibtex use spiebib.bst

\end{document}